\documentclass[showpacs,a4paper,12pt,prl,onecolumn]{revtex4}%
\usepackage{amsfonts}
\usepackage{amsmath}
\usepackage{amssymb}
\usepackage{graphicx}%
\providecommand{\U}[1]{\protect\rule{.1in}{.1in}}

\begin{document}
\title{Upper-bounded and sliced Jaynes- and anti-Jaynes-Cummings Hamiltonians and
Liouvillians in cavity quantum electrodynamics}
\author{W. Rosado$^{1}$, G. D. de Moraes Neto$^{1}$, F. O. Prado$^{2}$, and M. H. Y.
Moussa$^{1}$}
\affiliation{$^{1}$Instituto de F\'{\i}sica de S\~{a}o Carlos, Universidade de S\~{a}o
Paulo, Caixa Postal 369, 13560-970, S\~{a}o Carlos, S\~{a}o Paulo, Brazil}
\affiliation{$^{2}$Universidade Federal do ABC, Rua Santa Ad\'{e}lia 166, Santo Andr\'{e},
S\~{a}o Paulo 09210-170, Brazil}

\begin{abstract}
In this paper, we present a protocol to engineer upper-bounded and sliced
Jaynes-Cummings and anti-Jaynes-Cummings Hamiltonians in cavity quantum
electrodynamics. In the upper-bounded Hamiltonians, the atom-field interaction
is confined to a subspace of Fock states ranging from $\left\vert
0\right\rangle $ up to $\left\vert 4\right\rangle $, while in the sliced
interaction the Fock subspace ranges from $\left\vert M\right\rangle $ up to
$\left\vert M+4\right\rangle $. We also show how to build upper-bounded and
sliced Liouvillians irrespective of engineering Hamiltonians. The
upper-bounded and sliced Hamiltonians and Liouvillians can be used, among
other applications, to generate steady Fock states of a cavity mode and for
the implementation of a quantum-scissors device for optical state truncation.

\end{abstract}

\pacs{32.80.-t, 42.50.Ct, 42.50.Dv}
\maketitle

\section{Introduction}

The engineering of effective Hamiltonians has played an important role in
quantum information theory, where the design of interactions suited to
implementing quantum logical processes is a key ingredient \cite{CZ,Zheng,Nos}%
. Engineered interactions are also required for quantum simulation, where
controllable laboratory systems are assumed to simulate quantum phenomena that
share the same mathematical structure but are difficult to handle in the
laboratory \cite{QS}. Moreover, engineered interactions are a precondition for
the reservoir engineering technique, \cite{PCZ} where a target state is
protected. The strategy of switching off the reservoir and thus protecting any
quantum state, based on the engineering of a nonstationary quantum system
---where a time-dependent Hamiltonian must be appropriately prepared--- has
also been advanced \cite{Lucas}.

More recently, two schemes were presented to produce steady nonclassical
states within cavity QED \cite{Wilson} and trapped ions \cite{Rafael}, both
relying on suitably designed interactions. In the former case, selected
Jaynes-Cummings interactions are required where the two-level atom interacts
with only two neighboring Fock states of the cavity mode. These effective
interactions are further used for atomic reservoir engineering and steady Fock
states of the radiation field are produced. In the latter case, upper-,
lower-bounded and sliced Jaynes-Cummings (JC) and anti-Jaynes-Cummings (AJC)
Hamiltonians are required. The upper-bounded (lower-bounded) interaction acts
upon Fock subspaces ranging from $\left\vert 0\right\rangle $ to $\left\vert
M\right\rangle $ ($\left\vert N\right\rangle $ to$\ \infty$) and the sliced
one is confined to a Fock subspace ranging from $\left\vert M\right\rangle $
to $\left\vert N\right\rangle $, for any $M<N$. Whereas the upper-bounded
($ub$) JC or AJC interactions are shown to drive any initial state to a
quasi-steady Fock state $\left\vert N\right\rangle $, the sliced one is shown
to produce steady superpositions of Fock states confined to the sliced
subspace $\left\{  \left\vert N\right\rangle \text{,}\left\vert
N+1\right\rangle \right\}  $.

In this contribution, we present a strategy to engineer $ub$ and sliced JC and
AJC Hamiltonians in cavity QED which, as already shown in the context of
trapped ions \cite{Rafael}, can be used, among other purposes, to produce
steady Fock states and to implement a quantum-scissors device for optical
state truncation. We stress that the technique used to engineer $ub$ and
sliced interactions in cavity QED is completely different from that used with
trapped ions, where a suitable adjustment of the Lamb-Dicke parameter is
sufficient to bring about the desired interactions. In the other hand, as
should be clear below, in cavity QED the required interactions results from
appropriate adjustments in a sequence of Raman-type transitions. It is worth
emphasizing that our protocol can also be used in other platforms such as
circuit QED with fluxonium transitions \cite{CQED}.

Here we focus on the engineering of $ub$ Hamiltonians acting upon Fock
subspaces ranging from $\left\vert 0\right\rangle $ up to $\left\vert
4\right\rangle $. From these Hamiltonians we show how to derive master
equations with $ub$ Liouvillians via the atomic reservoir engineering scheme,
which are then employed to produce the steady Fock states $\left\vert
2\right\rangle $, $\left\vert 3\right\rangle $, and $\left\vert 4\right\rangle
$. Sliced Hamiltonians confined to Fock subspaces ranging from $\left\vert
M\right\rangle $ up to$\ \left\vert M+4\right\rangle $ are also derived and
used to generate sliced Liouvillians which produce steady superpositions of
Fock states confined to the associated sliced subspaces. Moreover, we present
a strategy to generate $ub$ and sliced master equations without needing first
to engineer $ub$ and sliced Hamiltonians, which can also be used to produce
steady Fock states. We advance that the present scheme can be applied to a
network of dissipative oscillators to produce steady entanglements in the
network corresponding to a steady Fock state in a given normal mode
\cite{Gentil}. Finally, we stress that the scheme used in the present
manuscript, as well as that in Ref. \cite{Wilson}, rely on a strategy to
engineer selective atom-field interactions presented in Ref. \cite{WilsonJOPB}%
. Other schemes to engineer selective transtions or quantum-scissors have also
been reported \cite{Others}.

It would be interesting at this point to recall an early theoretical proposal
to generate highly excited Fock states from an initial coherent state prepared
in the vibrational degrees of freedom of a trapped ion \cite{Vogel}. In that
scheme, the authors took only the relaxation mechanism of the electronic
states into account. Differently, in our proposal, we also consider the decay
of the cavity mode in which the steady Fock state is prepared; in fact, we
take advantage of this inevitable decay to produce our steady states, which
are achieved irrespectively of the initial state of the cavity mode. We also
note that various schemes have been presented to produce steady states in
cavity QED and trapped ion, other than those mentioned above \cite{PCZ,Wilson}%
. These protocols rely on reservoir engineering schemes \cite{re}, feedback
loops \cite{fl}, and quasi-local control \cite{qlc}. Nonequilibrium number
states up to $2$ photons have long been prepared in cavity QED \cite{Walther},
as in most suitable platforms, such as ion traps \cite{Leibfried}, and,
lately, in circuit QED \cite{Cleland}, where number states up to $6$ were
achieved. More recently, a quantum feedback technique was used to produce Fock
states with photon numbers up to $7$ and probability around $0.4$, in cavity
QED \cite{Haroche}.

\section{Upper-bounded and sliced JC and AJC interactions}

\subsection{Interactions confined to the Fock subspaces $\left\{  \left\vert
0\right\rangle ,\left\vert 1\right\rangle ,\left\vert 2\right\rangle \right\}
$ and $\left\{  \left\vert M\right\rangle ,\left\vert M+1\right\rangle
,\left\vert M+2\right\rangle \right\}  $}

We start by considering the engineering of $ub$ and sliced JC and AJC
interactions confined to the Fock subspaces $\left\{  \left\vert
0\right\rangle ,\left\vert 1\right\rangle ,\left\vert 2\right\rangle \right\}
$ and $\left\{  \left\vert M\right\rangle ,\left\vert M+1\right\rangle
,\left\vert M+2\right\rangle \right\}  $. To this end, we consider the atomic
level configuration sketched in Fig. 1(a), where the ground and excited states
$g$ and $e$ are coupled through Raman transitions to the auxiliary states $f$
and $h$. A cavity mode $\omega$ induces the transitions $g\leftrightarrow$
$f,h$ with strengths $\lambda_{1}$ and $\lambda_{2}$ and detunings $\Delta
_{1}$ and $\Delta_{2}$, while a laser field $\omega_{L}$ induces the
transitions $e\leftrightarrow$ $f,h$ with strengths $\Omega_{1}$ and
$\Omega_{2}$ and detunings $\tilde{\Delta}_{1}$ and $\tilde{\Delta}_{2}$. The
Hamiltonian describing the system in the interaction picture reads%
\begin{equation}
H=\lambda_{1}\sigma_{fg}a\operatorname*{e}\nolimits^{-i\Delta_{1}t}%
+\lambda_{2}\sigma_{hg}a\operatorname*{e}\nolimits^{i\Delta_{2}t}+\Omega
_{1}\sigma_{fe}\operatorname*{e}\nolimits^{-i\tilde{\Delta}_{1}t}+\Omega
_{2}\sigma_{he}\operatorname*{e}\nolimits^{i\tilde{\Delta}_{2}t}+H.c.,
\label{1}%
\end{equation}
where $\sigma_{rs}=\left\vert r\right\rangle \left\langle s\right\vert $, $r$
and $s$ labelling the atomic states involved, and we define the detunings
$\Delta_{1}=\omega-\omega_{f}$, $\Delta_{2}=\omega_{h}-\omega$, $\tilde
{\Delta}_{1}=\omega_{L}-\omega_{f}$, and $\tilde{\Delta}_{2}=\omega_{h}%
-\omega_{L}$. It is straightforward to show that, under the conditions
$\sqrt{\bar{n}+1}\lambda_{j}\ll\Delta_{j}$ ($\bar{n}$ being the mean
excitation of the field) and $\Omega_{j}\ll$ $\tilde{\Delta}_{j}$ ($j=1,2$),
we derive the effective interaction (\cite{James})
\[
H_{eff}=\chi a^{\dag}a\sigma_{gg}+\varpi\sigma_{ee}+%
{\textstyle\sum\nolimits_{j}}
\zeta_{j}\operatorname*{e}\nolimits^{i\theta_{j}t}\sigma_{ge}a^{\dagger}+H.c.
\]
where $\chi=\lambda_{1}^{2}/\Delta_{1}-\lambda_{2}^{2}/\Delta_{2}$ and
$\zeta_{j}=\left(  \lambda_{j}\Omega_{j}/2\right)  \left(  1/\Delta
_{j}+1/\tilde{\Delta}_{j}\right)  $ stand for off- and on-resonant atom-field
couplings, $\varpi=\Omega_{1}^{2}/\tilde{\Delta}_{1}-\Omega_{2}^{2}%
/\tilde{\Delta}_{2}$ is the frequency level shift due to the action of the
laser field, and $\theta_{j}=(-1)^{\delta_{1j}}\left(  \tilde{\Delta}%
_{j}-\Delta_{j}\right)  $ are convenient detunings required for the
engineering process. The unitary transformation $U=\exp\left[  -i\left(  \chi
a^{\dag}a\sigma_{gg}+\varpi\sigma_{ee}\right)  t\right]  $, takes $H_{eff}$
into the form
\[
H_{eff}=%
{\textstyle\sum\nolimits_{n}}
\left(  \zeta_{n}^{(1)}\operatorname*{e}\nolimits^{i\phi_{n}^{(1)}t}+\zeta
_{n}^{\left(  2\right)  }\operatorname*{e}\nolimits^{i\phi_{n}^{(2)}t}\right)
\sigma_{ge}\left\vert n+1\right\rangle \left\langle n\right\vert +H.c.,
\]
where $\phi_{n}^{\left(  j\right)  }=\xi_{n}+\theta_{j}$ and $\zeta_{n}%
^{(j)}=\sqrt{n+1}\zeta_{j}$, with $\xi_{n}=\left(  n+1\right)  \chi-\varpi$.
Finally, writing the Hamiltonian in the form%
\begin{align*}
-H_{eff}  &  =\zeta_{1}\operatorname*{e}\nolimits^{i\phi_{0}^{\left(
1\right)  }t}\left(  \left\vert 1\right\rangle \left\langle 0\right\vert
+\sqrt{2}\left\vert 2\right\rangle \left\langle 1\right\vert \operatorname*{e}%
\nolimits^{i\left(  \phi_{1}^{\left(  1\right)  }-\phi_{0}^{\left(  1\right)
}\right)  t}+...\right)  \sigma_{ge}\\
&  +\zeta_{2}\operatorname*{e}\nolimits^{i\phi_{1}^{\left(  2\right)  }%
t}\left(  \left\vert 1\right\rangle \left\langle 0\right\vert
\operatorname*{e}\nolimits^{i\left(  \phi_{0}^{\left(  2\right)  }-\phi
_{1}^{\left(  2\right)  }\right)  t}+\sqrt{2}\left\vert 2\right\rangle
\left\langle 1\right\vert +\sqrt{3}\left\vert 3\right\rangle \left\langle
2\right\vert \operatorname*{e}\nolimits^{i\left(  \phi_{2}^{\left(  2\right)
}-\phi_{1}^{\left(  2\right)  }\right)  t}+...\right)  \sigma_{ge}%
\end{align*}
we ready find that by adjusting the parameters such that
\begin{subequations}
\label{2l}%
\begin{align}
\phi_{0}^{\left(  1\right)  }  &  =\phi_{1}^{\left(  2\right)  }%
=0\text{,}\label{2la}\\
\left\vert \chi\right\vert  &  \gg\sqrt{2}\left\vert \zeta_{1}\right\vert
\text{,}\sqrt{3}\left\vert \zeta_{2}\right\vert \text{,} \label{2lb}%
\end{align}
we obtain the $ub$ Hamiltonian%

\end{subequations}
\begin{equation}
H_{ub}=\zeta_{1}\sigma_{ge}A_{ub}^{\dagger}+H.c., \label{3}%
\end{equation}
with the field operator
\[
A_{ub}^{\dagger}=\left\vert 1\right\rangle \left\langle 0\right\vert +\left(
\sqrt{2}\zeta_{2}/\zeta_{1}\right)  \left\vert 2\right\rangle \left\langle
1\right\vert \text{.}%
\]

We have thus engineered the $ub$ JC Hamiltonian confined to the Fock
subspace\ $\left\{  \left\vert 0\right\rangle ,\left\vert 1\right\rangle
,\left\vert 2\right\rangle \right\}  $. To achieve a sliced JC Hamiltonian
confined to the Fock subspace\ $\left\{  \left\vert M\right\rangle ,\left\vert
M+1\right\rangle ,\left\vert M+2\right\rangle \right\}  $, we redefine the
conditions in Eq. (\ref{2l}) to
\begin{align*}
\phi_{M}^{\left(  1\right)  }  &  =\phi_{M+1}^{\left(  2\right)  }=0\text{,}\\
\left\vert \chi\right\vert  &  \gg\left\vert \zeta_{M+1}^{(1)}\right\vert
\text{,}\left\vert \zeta_{M+2}^{\left(  2\right)  }\right\vert \text{,}%
\end{align*}
leading us to the sliced Hamiltonian
\begin{equation}
H_{s}=\zeta_{M}^{\left(  1\right)  }\sigma_{ge}A_{s}^{\dagger}+H.c., \label{6}%
\end{equation}
with the field operator
\[
A_{s}^{\dagger}=\left\vert M+1\right\rangle \left\langle M\right\vert +\left(
\zeta_{M+1}^{\left(  2\right)  }/\zeta_{M}^{\left(  1\right)  }\right)
\left\vert M+2\right\rangle \left\langle M+1\right\vert \text{.}%
\]

\begin{figure}
\includegraphics[width=0.7\columnwidth]{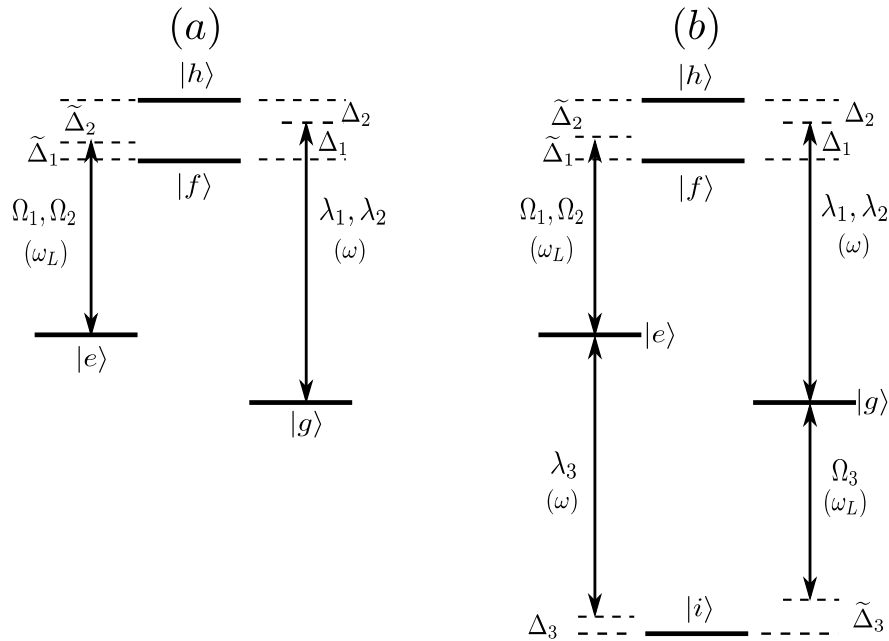}%
\caption{(a) Four-level atomic configuration to engineer the $ub$ and sliced JC
Hamiltonians acting upon Fock subspaces ranging from $\left\vert
0\right\rangle $ to $\left\vert 2\right\rangle $ and $\left\vert
M\right\rangle $ to $\left\vert M+2\right\rangle $, and (b) Five-level atomic
configuration to engineer the $ub$ and sliced JC Hamiltonians acting upon Fock
subspaces ranging from $\left\vert 0\right\rangle $ to $\left\vert
3\right\rangle $ and $\left\vert M\right\rangle $ to $\left\vert
M+3\right\rangle $.}%

\end{figure}
It is worth noting that both the $ub$ and sliced AJC Hamiltonians follow from
the same atomic configuration in Fig. 1(a), but with the cavity mode and the
laser field tuned to the reversed transitions $e\leftrightarrow$ $f,h$ and
$g\leftrightarrow$ $f,h$, respectively. We also observe that, instead of the
four-level configuration presented in Fig. 1(a), a rhombus-like level
configuration can also be used to generate Hamiltonians (\ref{3}) and
(\ref{6}). This latter configuration avoids the required proximity of levels
$f$ and $h$ in Fig. 1, maintaining, however, all the conditions and relations
exactly as derived above. Finally, we observe that the AJC interactions
confined to the Fock subspaces $\left\{  \left\vert 0\right\rangle ,\left\vert
1\right\rangle ,\left\vert 2\right\rangle \right\}  $ and $\left\{  \left\vert
M\right\rangle ,\left\vert M+1\right\rangle ,\left\vert M+2\right\rangle
\right\}  $ can be generated by interchanging the transitions driven by the
cavity mode and the laser, tuning the quantum (classical) field to drive the
transition $e$ ($g$) $\leftrightarrow$ $f,h$ instead of $g$ ($e$)
$\leftrightarrow$ $f,h$.

\subsection{Interactions confined to the Fock subspaces $\left\{  \left\vert
0\right\rangle ,\left\vert 1\right\rangle ,\left\vert 2\right\rangle
,\left\vert 3\right\rangle \right\}  $ and $\left\{  \left\vert M\right\rangle
,\left\vert M+1\right\rangle ,\left\vert M+2\right\rangle ,\left\vert
M+3\right\rangle \right\}  $}

To engineer the $ub$ and sliced Hamiltonians confined to Fock subspaces
$\left\{  \left\vert 0\right\rangle ,\left\vert 1\right\rangle ,\left\vert
2\right\rangle ,\left\vert 3\right\rangle \right\}  $ and $\left\{  \left\vert
M\right\rangle ,\left\vert M+1\right\rangle ,\left\vert M+2\right\rangle
,\left\vert M+3\right\rangle \right\}  $, we must add another level
$\left\vert i\right\rangle $ to the atomic configuration sketched in Fig.
1(a), as shown in Fig. 1(b). Now, the cavity mode (laser field) also induces
the Raman transition $e\leftrightarrow i$ ($g\leftrightarrow i$) to the
additional auxiliary state with coupling strength $\lambda_{3}$ ($\Omega_{3}$)
and detuning $\Delta_{3}$ ($\tilde{\Delta}_{3}$). The Hamiltonian describing
the system in the interaction picture reads%

\begin{align}
H  &  =\lambda_{1}\sigma_{fg}a\operatorname*{e}\nolimits^{-i\Delta_{1}%
t}+\lambda_{2}\sigma_{hg}a\operatorname*{e}\nolimits^{i\Delta_{2}t}%
+\lambda_{3}\sigma_{hg}a\operatorname*{e}\nolimits^{i\Delta_{3}t}\nonumber\\
&  +\Omega_{1}\sigma_{fe}\operatorname*{e}\nolimits^{-i\tilde{\Delta}_{1}%
t}+\Omega_{2}\sigma_{he}\operatorname*{e}\nolimits^{i\tilde{\Delta}_{2}%
t}+\Omega_{2}\sigma_{he}\operatorname*{e}\nolimits^{i\tilde{\Delta}_{3}%
t}+H.c., \label{8}%
\end{align}
where $\Delta_{3}=\omega_{e}+\omega_{i}-\omega$, $\tilde{\Delta}_{3}%
=\omega_{i}-\omega_{L}$, and all the other parameters are those defined above.
By analogy with the former reasoning used to obtain Hamiltonians (\ref{3}) and
(\ref{6}), we next consider the conditions $\sqrt{\bar{n}+1}\lambda_{j}%
\ll\tilde{\Delta}_{j}$, $\Omega_{j}\ll$ $\Delta_{j}$ ($j=1,2,3$), and
\begin{subequations}
\label{8l}%
\begin{align}
\Phi_{0}^{\left(  1\right)  }  &  =\Phi_{1}^{\left(  2\right)  }=\Phi
_{2}^{\left(  3\right)  }=0\text{,}\label{8la}\\
\left\vert \chi-\tilde{\chi}\right\vert  &  \gg\sqrt{2}\left\vert \zeta
_{1}\right\vert \text{,}\sqrt{3}\left\vert \zeta_{2}\right\vert \text{,}%
2\left\vert \zeta_{3}\right\vert \text{,} \label{8lb}%
\end{align}
where $\Phi_{n}^{\left(  j\right)  }=\Xi_{n}+\theta_{j\text{,}}$ $\Xi_{n}%
=\xi_{n}-\left(  n+1\right)  \tilde{\chi}+\Omega_{3}^{2}/\tilde{\Delta}_{3}$,
$\tilde{\chi}=\lambda_{3}^{2}/\Delta_{3}$, and $\xi_{n}$, $\theta_{j\text{,}}%
$, and $\zeta_{j}$ are as defined above, now with $j=1,2,3$. When the
auxiliary level $i$ is disregarded, the uppercase parameters ($\Phi
_{n}^{\left(  j\right)  }$,$\Xi_{n}$) recover those in lowercase ($\phi
_{n}^{\left(  j\right)  }$,$\xi_{n}$) for $j=1,2$. Under the above conditions,
we derive the $ub$ Hamiltonian%

\end{subequations}
\begin{equation}
H_{ub}=\zeta_{1}\sigma_{ge}B_{ub}^{\dagger}+H.c., \label{9}%
\end{equation}
with the field operator
\begin{equation}
A_{ub}^{\dagger}=\left\vert 1\right\rangle \left\langle 0\right\vert +\left(
\sqrt{2}\zeta_{2}/\zeta_{1}\right)  \left\vert 2\right\rangle \left\langle
1\right\vert +\left(  \sqrt{3}\zeta_{3}/\zeta_{1}\right)  \left\vert
3\right\rangle \left\langle 2\right\vert \text{.} \label{10}%
\end{equation}

Again, we achieve a sliced JC Hamiltonian confined to the Fock
subspace\ $\left\{  \left\vert M\right\rangle ,\left\vert M+1\right\rangle
,\left\vert M+2\right\rangle ,\left\vert M+3\right\rangle \right\}  $ by
redefining the conditions in Eq. (\ref{8l}) to
\begin{align*}
\Phi_{M}^{\left(  1\right)  }  &  =\Phi_{M+1}^{\left(  2\right)  }=\Phi
_{M+2}^{\left(  3\right)  }=0\text{,}\\
\left\vert \chi\right\vert  &  \gg\left\vert \zeta_{M+1}^{(1)}\right\vert
\text{,}\left\vert \zeta_{M+2}^{\left(  2\right)  }\right\vert \text{,}%
\left\vert \zeta_{M+3}^{\left(  3\right)  }\right\vert \text{,}%
\end{align*}
leading us to the sliced interaction
\begin{equation}
H_{s}=\zeta_{M}^{\left(  1\right)  }\sigma_{ge}B_{s}^{\dagger}+H.c.,
\label{11}%
\end{equation}
with the field operator
\[
A_{s}^{\dagger}=\left\vert M+1\right\rangle \left\langle M\right\vert +\left(
\zeta_{M+1}^{\left(  2\right)  }/\zeta_{M}^{\left(  1\right)  }\right)
\left\vert M+2\right\rangle \left\langle M+1\right\vert +\left(  \zeta
_{M+2}^{\left(  3\right)  }/\zeta_{M}^{\left(  1\right)  }\right)  \left\vert
M+3\right\rangle \left\langle M+2\right\vert \text{.}%
\]

To generate the AJC interactions confined to the Fock subspaces $\left\{
\left\vert 0\right\rangle ,\left\vert 1\right\rangle ,\left\vert
2\right\rangle \right\}  $ and $\left\{  \left\vert M\right\rangle ,\left\vert
M+1\right\rangle ,\left\vert M+2\right\rangle \right\}  $, we have again to
interchange the transitions $g,e\leftrightarrow i$ driven by the cavity mode
and the laser field. Finally, we observe that, by inserting an additional
auxiliary level close to $i$, we can obtain $ub$ and sliced interactions
confined to the Fock subspaces ranging from $\left\vert 0\right\rangle $ to
$\left\vert 4\right\rangle $ and $\left\vert M\right\rangle $ to $\left\vert
M+4\right\rangle $.

\section{Validity of the engineered interactions}

We next analyze the validity of the engineered $ub$ and sliced Hamiltonians
described by Eqs. \ref{3}, \ref{6}, \ref{9}, and \ref{11}. To this end, we
compare numerically the sinusoidal Rabi oscillations of the cavity mode
population coming from these effective interactions with those derived from
the complete Hamiltonians in Eqs. \ref{1} and \ref{8}.

\subsection{Interactions confined to the Fock subspaces $\left\{  \left\vert
0\right\rangle ,\left\vert 1\right\rangle ,\left\vert 2\right\rangle \right\}
$ and $\left\{  \left\vert M\right\rangle ,\left\vert M+1\right\rangle
,\left\vert M+2\right\rangle \right\}  $}

We start by analyzing the validity of the engineered $ub$ interaction given by
Eq. (\ref{3}), confined to the Fock subspaces $\left\{  \left\vert
0\right\rangle ,\left\vert 1\right\rangle ,\left\vert 2\right\rangle \right\}
$. Assuming the initial state $\left(  \left\vert 0\right\rangle +\left\vert
2\right\rangle \right)  \otimes\left(  \left\vert g\right\rangle +\left\vert
e\right\rangle \right)  /2$, in Fig. 2(a) we plot, against the interaction
parameter $\zeta_{1}t$, the probabilities $\mathcal{P}_{0}$, $\mathcal{P}_{1}%
$, and $\mathcal{P}_{2}$ of measuring the cavity mode in Fock states
$\left\vert 0\right\rangle $, $\left\vert 1\right\rangle $, and $\left\vert
2\right\rangle $. We have assumed typical coupling strengths, $\lambda
_{1}=\lambda_{2}=5\times10^{5}$ Hz \cite{RMP}, and set the detunings to
$\Delta_{1}=2\Delta_{2}=10\lambda_{1}$ and $\theta_{1}=\theta_{2}%
/2=\lambda_{1}/10$, to obtain $\tilde{\Delta}_{1}=99\tilde{\Delta}%
_{2}/52=9.9\lambda_{1}$. Finally, with $\Omega_{1}=2\sqrt{2}\Omega_{2}=$
$\lambda_{1}/5\sqrt{2}$, we get $\zeta_{1}=\sqrt{2}\zeta_{2}=1.7\times10^{3}$
Hz and, consequently, the $ub$ operator:\textbf{ }$A_{ub}^{\dagger}=\left\vert
1\right\rangle \left\langle 0\right\vert +\left\vert 2\right\rangle
\left\langle 1\right\vert $\textbf{.} These parameters seem quite suitable for
the derivation of the effective interaction, as confirmed by the good
agreement in Fig. 2(a) between the light curves, derived from the full
Hamiltonian (\ref{1}), and the dark ones computed from the engineered
interaction (\ref{3}), given by $\mathcal{P}_{0}=\mathcal{P}_{2}=\left[
1+\cos^{2}\left(  \zeta_{1}t\right)  \right]  /4$ and $\mathcal{P}_{1}%
=\sin^{2}\left(  \zeta_{1}t\right)  /2$. While $\mathcal{P}_{0}$ and
$\mathcal{P}_{1}$ are exposed in the foreground, $\mathcal{P}_{2}$ is shown in
the inset. The probability $\mathcal{P}_{3}$, which is evidently null when
computed from the effective interaction, oscillates near zero for the full
Hamiltonian, as shown by the dashed curve.

To analyze the validity of the engineered sliced interaction in Eq. (\ref{6}),
we assume $M=3$ to confine it to the Fock subspace $\left\{  \left\vert
3\right\rangle ,\left\vert 4\right\rangle ,\left\vert 5\right\rangle \right\}
$. We assume the initial state $\left(  \left\vert 3\right\rangle +\left\vert
5\right\rangle \right)  \otimes\left(  \left\vert g\right\rangle +\left\vert
e\right\rangle \right)  /2$ to plot the probabilities $\mathcal{P}_{3}$,
$\mathcal{P}_{4}$, and $\mathcal{P}_{5}$, in Fig. 2(b), against the
interaction parameter $\zeta_{3}^{\left(  1\right)  }t$ (with $\zeta
_{3}^{\left(  1\right)  }=2\zeta_{1}$), of measuring the cavity mode in Fock
states $\left\vert 3\right\rangle $, $\left\vert 4\right\rangle $, and
$\left\vert 5\right\rangle $. With typical coupling strengths, $\lambda
_{1}=\lambda_{2}=5\times10^{5}$ Hz, and setting $\Delta_{1}=2\Delta
_{2}=20\lambda_{1}$ and $\theta_{1}=4\theta_{2}/5=\lambda_{1}/5$, we obtain
$\tilde{\Delta}_{1}=79.2\tilde{\Delta}_{2}/41=19.8\lambda_{1}$. Moreover, with
laser-field strengths $\Omega_{1}=\sqrt{5}\Omega_{2}=$ $\lambda_{1}/5\sqrt{5}%
$, we obtain $\zeta_{1}=\sqrt{2}\zeta_{2}=6\times10^{2}$ Hz and, consequently:
$A_{s}^{\dagger}=\left\vert 4\right\rangle \left\langle 3\right\vert
+\left\vert 5\right\rangle \left\langle 4\right\vert $. Again we observe good
agreement between the light curves, numerically computed from the full
Hamiltonian (\ref{1}), and the dark ones, analytically computed from the
engineered interaction (\ref{6}), given by $\mathcal{P}_{3}=\mathcal{P}%
_{5}=\left[  1+\cos^{2}\left(  \zeta_{3}^{\left(  1\right)  }t\right)
\right]  /4$ and $\mathcal{P}_{4}=\sin^{2}\left(  \zeta_{3}^{\left(  1\right)
}t\right)  /2$. While $\mathcal{P}_{3}$ and $\mathcal{P}_{4}$ are exposed in
the foreground, $\mathcal{P}_{5}$ is shown in the inset. The probabilities
$\mathcal{P}_{2}$ and $\mathcal{P}_{6}$, which are null for the effective
interaction, oscillate near zero for the full Hamiltonian, as shown by the
dashed curves in the foreground and the inset, respectively.
\begin{figure}
\includegraphics[width=1.0\columnwidth]{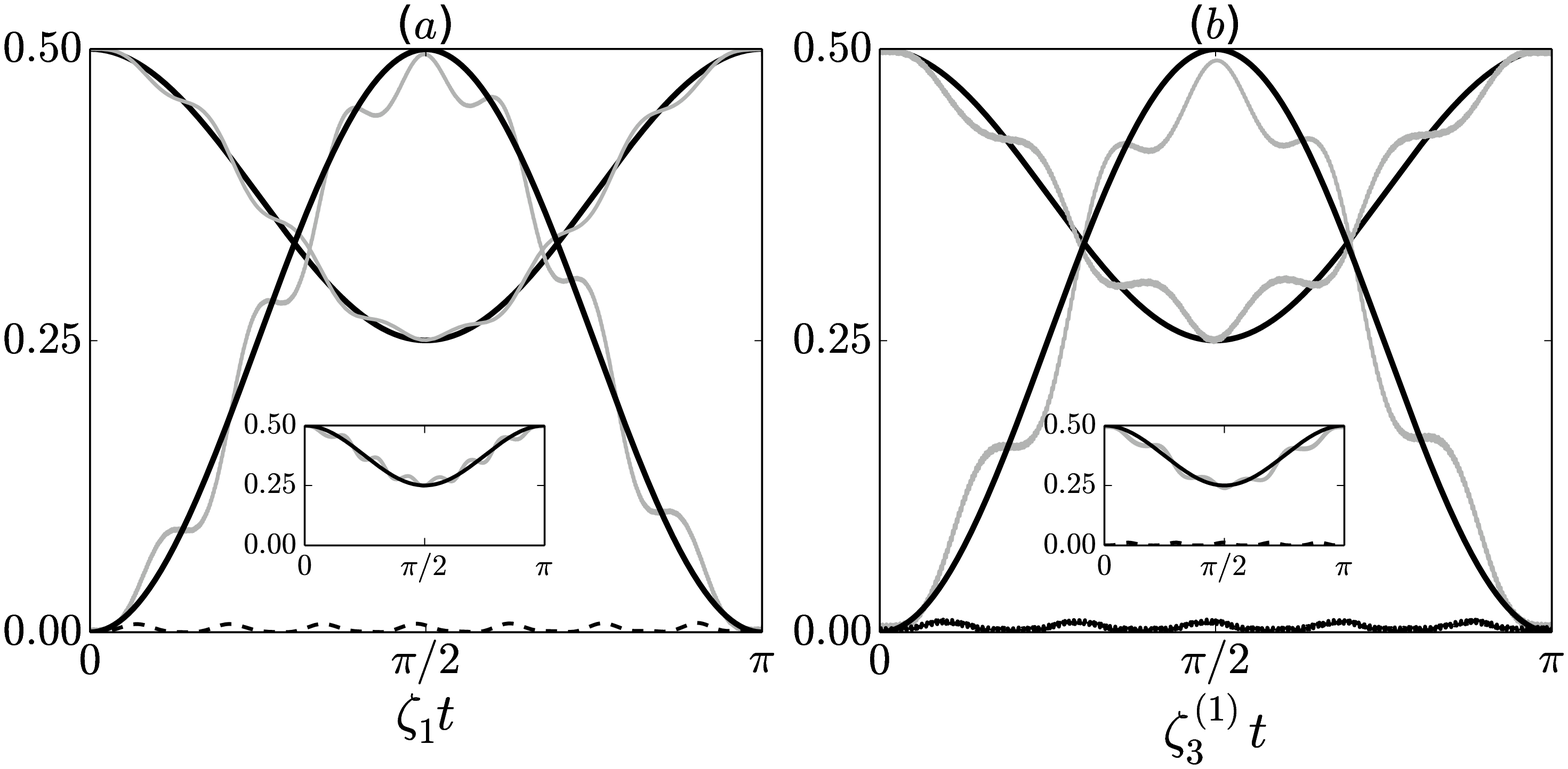}%
\caption{(a) Probabilities $\mathcal{P}_{0}$ to $\mathcal{P}_{3}$ of measuring
the cavity mode in the Fock states $\left\vert 0\right\rangle $ to $\left\vert
3\right\rangle $, computed from the engineered (\ref{3}) and the full
(\ref{1}) Hamiltonians, shown by dark and light lines, respectively.
$\mathcal{P}_{0}$ and $\mathcal{P}_{1}$ are exposed in the foreground while
$\mathcal{P}_{2}$ is shown in the inset. $\mathcal{P}_{3}$, which is null for
the effective interaction, oscillates near zero for the full Hamiltonian, as
shown by the dashed curve. (b) Probabilities $\mathcal{P}_{2}$ to
$\mathcal{P}_{6}$ of measuring the cavity mode in the Fock states $\left\vert
3\right\rangle $ to $\left\vert 5\right\rangle $, computed from the engineered
(\ref{6}) and the full (\ref{1}) Hamiltonians, shown by dark and light lines,
respectively. $\mathcal{P}_{3}$ and $\mathcal{P}_{4}$ are exposed in the
foreground while $\mathcal{P}_{5}$ is shown in the inset. $\mathcal{P}_{2}$
and $\mathcal{P}_{6}$, which are null for the effective interaction,
oscillates near zero for the full Hamiltonian as shown by the dashed curves in
the foreground and the inset, respectively.}%
\end{figure}
\subsection{Interactions confined to the Fock subspaces $\left\{  \left\vert
0\right\rangle ,\left\vert 1\right\rangle ,\left\vert 2\right\rangle
,\left\vert 3\right\rangle \right\}  $ and $\left\{  \left\vert M\right\rangle
,\left\vert M+1\right\rangle ,\left\vert M+2\right\rangle ,\left\vert
M+3\right\rangle \right\}  $}

Next, analyzing the validity of the engineered $ub$ interaction (\ref{9}),
confined to the Fock subspace $\left\{  \left\vert 0\right\rangle ,\left\vert
1\right\rangle ,\left\vert 2\right\rangle ,\left\vert 3\right\rangle \right\}
$, we assume the initial state $\left(  \left\vert 1\right\rangle +\left\vert
3\right\rangle \right)  \otimes\left(  \left\vert g\right\rangle +\left\vert
e\right\rangle \right)  /2$, so as to plot the probabilities $\mathcal{P}_{0}$
to $\mathcal{P}_{3}$ against the interaction parameter $\zeta_{1}t$, in Fig.
3(a). The typical strengths, $\lambda_{1}=\lambda_{2}=\lambda_{3}%
=5\times10^{5}$ Hz, and the detunings set at $\Delta_{1}=\Delta_{2}%
/2=\Delta_{3}=10\lambda_{1}$ and $\theta_{1}=\theta_{2}/2=\theta_{3}%
/3=\lambda_{1}/20$, lead to $\tilde{\Delta}_{1}=199\tilde{\Delta}%
_{2}/402=199\tilde{\Delta}_{3}/203=10.15\lambda_{1}$. Finally, with
$\Omega_{1}=\Omega_{2}/\sqrt{2}=\sqrt{3}\Omega_{3}=\lambda_{1}/20$, we obtain
$\zeta_{1}=\sqrt{2}\zeta_{2}=\sqrt{3}\zeta_{3}=1.77\times10^{3}$ Hz and,
consequently:\textbf{ }$A_{ub}^{\dagger}=\left\vert 1\right\rangle
\left\langle 0\right\vert +\left\vert 2\right\rangle \left\langle 1\right\vert
+\left\vert 3\right\rangle \left\langle 2\right\vert $\textbf{. }As in the
figures analyzed above, the light curves, derived from the full Hamiltonian,
are in good agreement with the dark ones arising from the engineered
interaction (\ref{9}) and given by $\mathcal{P}_{0}=\mathcal{P}_{2}/2=\sin
^{2}\left(  \zeta_{1}t\right)  /4$ and $\mathcal{P}_{1}=2\mathcal{P}%
_{3}-1/2=\cos^{2}\left(  \zeta_{1}t\right)  /2$. While $\mathcal{P}_{0}$ and
$\mathcal{P}_{1}$ are in the foreground, $\mathcal{P}_{2}$ and $\mathcal{P}%
_{3}$ are in the inset. The probability $\mathcal{P}_{4}$, which is null when
computed from the effective interaction, oscillates near zero for the full
Hamiltonian, as shown by the dashed curve.

Regarding the validity of the engineered sliced interaction (\ref{11}), we
assume $M=3$ to confine it to the Fock subspace $\left\{  \left\vert
3\right\rangle ,\left\vert 4\right\rangle ,\left\vert 5\right\rangle
,\left\vert 6\right\rangle \right\}  $. Assuming the initial state $\left(
\left\vert 3\right\rangle +\left\vert 6\right\rangle \right)  \otimes\left(
\left\vert g\right\rangle +\left\vert e\right\rangle \right)  /2$, we plot the
probabilities $\mathcal{P}_{3}$ to $\mathcal{P}_{6}$ against the interaction
parameter $\zeta_{3}^{\left(  1\right)  }t$ (with $\zeta_{3}^{\left(
1\right)  }=2\zeta_{1}$), in Fig. 3(b). With the typical coupling strengths,
$\lambda_{1}=\lambda_{2}=\lambda_{3}=5\times10^{5}$ Hz and the detunings set
at $\Delta_{1}=\Delta_{2}/2=\Delta_{3}=20\lambda_{1}$ and $\theta_{1}%
=4\theta_{2}/5=2\theta_{3}/3=\lambda_{1}/10$, we have $\tilde{\Delta}%
_{1}=159.2\tilde{\Delta}_{2}/321=199\tilde{\Delta}_{3}/201.5=20.15\lambda_{1}%
$. Moreover, with the laser-field strengths $\Omega_{1}=\sqrt{5}\Omega
_{2}/4=\sqrt{6}\Omega_{3}/2=\lambda_{1}/20\sqrt{5}$, we have $\zeta_{1}%
=\sqrt{5}\zeta_{2}/2=\sqrt{6}\zeta_{3}/2=1.1\times10^{3}$ Hz and,
consequently: $A_{ub}^{\dagger}=\left\vert 4\right\rangle \left\langle
3\right\vert +\left\vert 5\right\rangle \left\langle 4\right\vert +\left\vert
6\right\rangle \left\langle 5\right\vert $. We observe good agreement between
the light curves, computed from the full Hamiltonian, and the dark ones,
calculated from the engineered interaction (\ref{11}), given by $\mathcal{P}%
_{3}=\mathcal{P}_{6}=\left[  1+\cos^{2}\left(  \zeta_{3}^{\left(  1\right)
}t\right)  \right]  /4$ and $\mathcal{P}_{4}=\mathcal{P}_{5}=\sin^{2}\left(
\zeta_{3}^{\left(  1\right)  }t\right)  /4$. While $\mathcal{P}_{3}$ and
$\mathcal{P}_{4}$ are exposed in the foreground, $\mathcal{P}_{5}$ and
$\mathcal{P}_{6}$ are shown in the inset. The probabilities $\mathcal{P}_{2}$
and $\mathcal{P}_{7}$, which are null for the effective interaction, oscillate
near zero for the full Hamiltonian, as shown by the dashed curves in the
foreground and the inset, respectively.
\begin{figure}
\includegraphics[width=1.0\columnwidth]{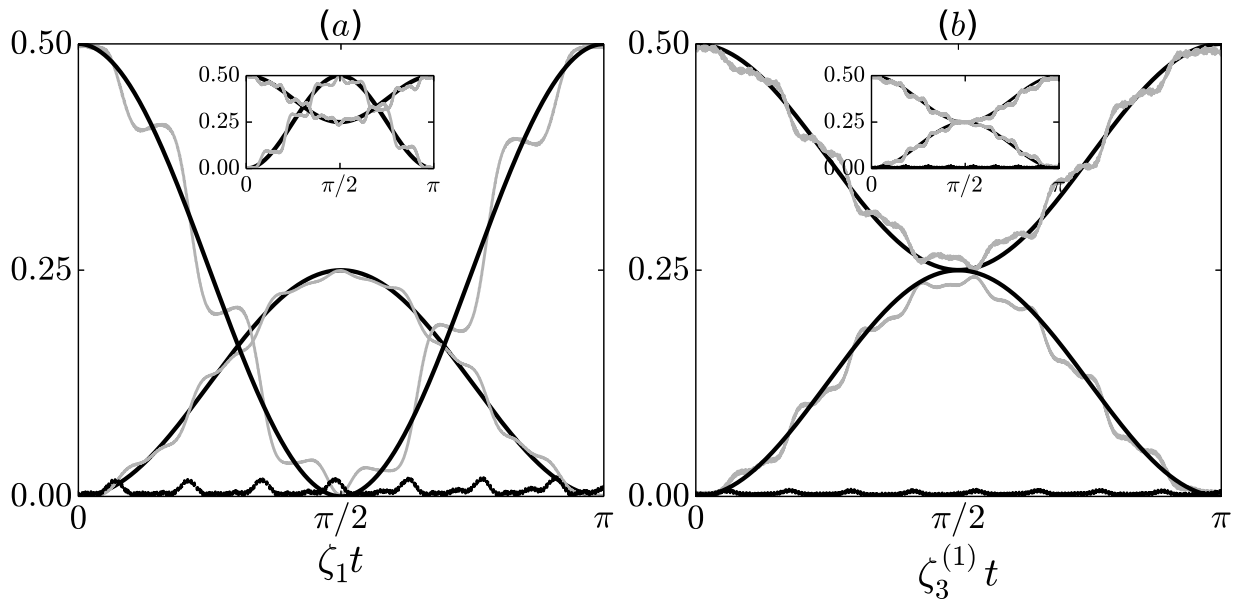}%
\caption{(a) Probabilities $\mathcal{P}_{0}$ to $\mathcal{P}_{4}$ of measuring
the cavity mode in the Fock states $\left\vert 0\right\rangle $ to $\left\vert
4\right\rangle $, computed from the engineered (\ref{9}) and the full
(\ref{1}) Hamiltonians, shown by dark and light lines, respectively.
$\mathcal{P}_{0}$ and $\mathcal{P}_{1}$ are exposed in the foreground while
$\mathcal{P}_{2}$ and $\mathcal{P}_{3}$ are in the inset. $\mathcal{P}_{4}$,
which is null for the effective interaction, oscillate near zero for the full
Hamiltonian, as shown by the dashed curve. (b) Probabilities $\mathcal{P}_{2}$
to $\mathcal{P}_{7}$ of measuring the cavity mode in the Fock states
$\left\vert 2\right\rangle $ to $\left\vert 7\right\rangle $, computed from
the engineered (\ref{11}) and the full (\ref{1}) Hamiltonians, shown by dark
and light lines, respectively. $\mathcal{P}_{3}$ and $\mathcal{P}_{4}$ are
exposed in the foreground while $\mathcal{P}_{5}$ and $\mathcal{P}_{6}$ are in
the inset. $\mathcal{P}_{2}$ and $\mathcal{P}_{7}$, which are null for the
effective interaction, oscillate near zero for the full Hamiltonian, as shown
by the dashed curves in the foreground and the inset, respectively.}%
\label{figw}
\end{figure}
\section{Atomic reservoir engineering}

Turning to the applications of the above-derived $ub$ Hamiltonians, we next
present a method to protect the Fock state $\left\vert 3\right\rangle $, which
relies on an engineered atomic reservoir \cite{ABR,BR}. A bunch of atoms are
required to interact with the cavity mode, one at a time, through the
engineered $ub$ Hamiltonian (\ref{9}). Assuming that all the atoms are
prepared in the excited state before interacting with the cavity mode under
the weak coupling regime $\zeta_{1}\tau\ll1$, $\tau$ being the average transit
time of the atoms through the cavity, it is straightforward to obtain the
master equation
\begin{equation}
\dot{\rho}=\frac{\Gamma}{2}\left(  2A_{ub}^{\dagger}\rho A_{ub}\mathbf{-}%
A_{ub}A_{ub}^{\dagger}\rho-\rho A_{ub}A_{ub}^{\dagger}\right)  +\mathcal{L}%
\rho, \label{Eq7}%
\end{equation}
where $A_{ub}$ is given by Eq. (\ref{10}) and the effective damping rate
$\Gamma=r\left(  \zeta_{1}\tau\right)  ^{2}$, $r$ being the arrival rate of
atoms. The Lindblad form
\begin{align}
\mathcal{L}\rho &  =\frac{\gamma}{2}\left(  1+\bar{n}\right)  \left(  2a\rho
a^{\dag}-\rho a^{\dag}a-a^{\dag}a\rho\right) \nonumber\\
&  {\small +}\frac{\gamma}{2}\bar{n}\left(  2a^{\dag}\rho a-\rho aa^{\dag
}-aa^{\dag}\rho\right)  , \label{Eq8}%
\end{align}
describes the effect of the natural environment, at temperature $T=\hbar
\omega/k_{B}\ln\left[  \left(  1+\bar{n}\right)  /\bar{n}\right]  $, on the
cavity mode, with damping rate $\gamma$.

It is not difficult to conclude that under the condition $\Gamma\gg\gamma$,
i.e., $r\gg\gamma/\left(  \zeta_{1}\tau\right)  ^{2}$, the cavity mode
---whatever its initial state--- is asymptotically driven to a quasi-steady
Fock state $\left\vert 3\right\rangle $, which is the only eigenstate of
$A^{\dagger}$ with null eigenvalue ($A_{ub}^{\dagger}\left\vert 3\right\rangle
=0$). This occurs because, when $\Gamma\gg\gamma$, the engineered contribution
to $\dot{\rho}$, confined to the subspace ranging from $\left\vert
0\right\rangle $ to $\left\vert 3\right\rangle $, prevails over the action of
the thermal environment described by Eq. (\ref{Eq8}).

Assuming the same parameters as in Fig. 3(a), such that $\zeta_{1}%
=1.77\times10^{3}$ Hz, with $\tau=1/r=2\times10^{-4}$ s, we obtain the
engineered decay rate $\Gamma=63\gamma$, for a high-$Q$ cavity ($\gamma\sim
10$Hz). In Fig. 4, starting with a thermal state with $\bar{n}=0.05$, we
present the evolution of the fidelity $\mathcal{F}_{3}(t)=$
$\operatorname*{Tr}\left\vert 3\right\rangle \left\langle 3\right\vert
\rho(t)$ against $\gamma t$, showing that the Liouvillian (\ref{Eq7}),
engineered through the $ub$ Hamiltonian (\ref{9}), leads to the steady Fock
state $\left\vert 3\right\rangle $ with fidelity around $0.92$. In the inset,
we plot Mandel's $Q$ factor \cite{WM,SZ}, which starts from around $0.05$ and
goes to $-0.96$, indicating that the final steady state indeed approaches a
Fock state.

We finally remark that, although the coupling strength of the engineered $ub$
Hamiltonian is small in comparison with typical Rabi frequencies in microwave
cavity QED, this is evidently not a limiting factor for the present proposal
for the construction of stationary Fock states. Moreover, this technique to
produce stationary Fock states also applies to the case of state $\left\vert
2\right\rangle $, reached from a Liouvillian engineered from interaction
(\ref{3}), and state $\left\vert 4\right\rangle $, reached from an engineered
interaction confined to the Fock subspace $\left\{  \left\vert 0\right\rangle
,\left\vert 1\right\rangle ,\left\vert 2\right\rangle ,\left\vert
3\right\rangle ,\left\vert 4\right\rangle \right\}  $.
\begin{figure}
\includegraphics[width=0.7\columnwidth]{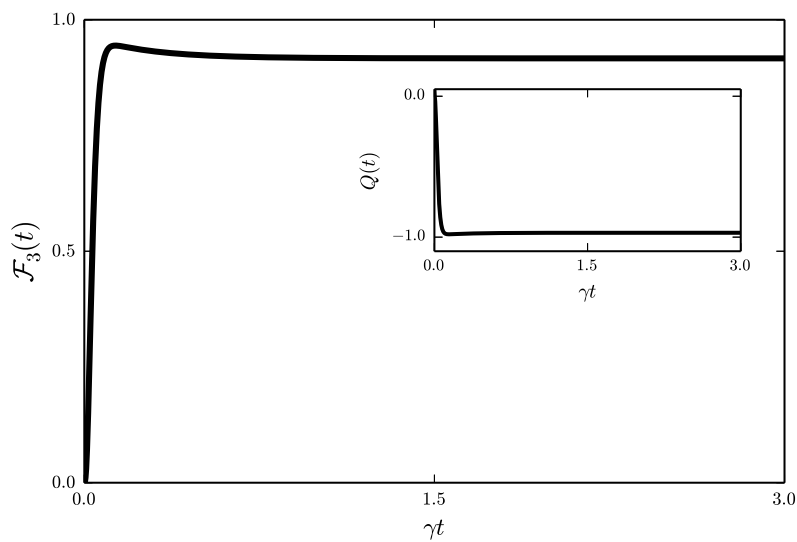}%
\caption{Evolution of the fidelity $\mathcal{F}_{3}(t)=$ $\operatorname*{Tr}%
\left\vert 3\right\rangle \left\langle 3\right\vert \rho(t)$, governed by the
upper-bound Hamiltonian (\ref{3}), showing that a steady Fock state
$\left\vert 3\right\rangle $ is reached with a significant fidelity, around
$0.92$, from a thermal state. The evolution of Mandel's $Q$ factor is shown in
the inset, indicating that the final steady state indeed approaches a Fock state.}%

\end{figure}
\section{Upper-bounded Liouvillians independent of upper-bounded Hamiltonians}

We now present a strategy to generate $ub$ Liouvillians in cavity QED which
does not rely on an engineered $ub$ Hamiltonian. To this end, we follow the
line of reasoning in Ref. \cite{Wilson}, where selective Liouvillians are
reached by combining engineered selective Hamiltonians ---governing the
interaction of two-level atoms with only two Fock field states
\cite{WilsonJOPB}--- and engineered atomic reservoirs \cite{ABR,BR}.
\begin{figure}
\includegraphics[width=0.4\columnwidth]{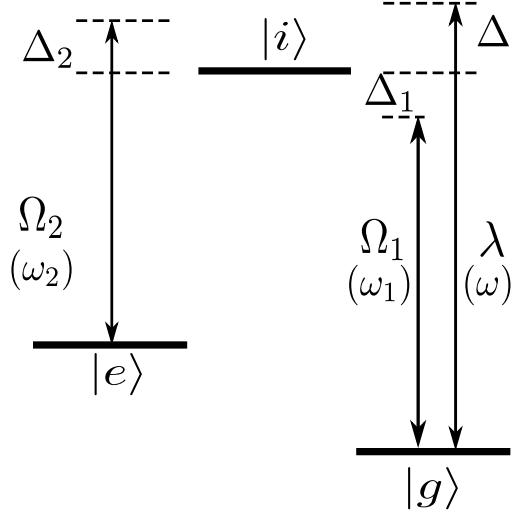}%
\caption{Atomic level configurations to engineer selective Hamiltonians.}%

\end{figure}
For the sake of clarity, we briefly revisit the ideas developed in Ref.
\cite{Wilson}, remembering that the engineering of atomic reservoirs requires
a beam of atoms to cross the cavity, interacting one at a time with the cavity
mode under the weak coupling regime. Each atom must interact simultaneously
with the cavity mode and a pair of laser beams, via the level configuration in
Fig. 5, where an auxiliary level $i$ is considered apart from the ground and
excited states $g$ and $e$. The cavity mode ($\omega$) is used to promote a
Raman-type transition $g\leftrightarrow e$, helped by the laser beams,
$\omega_{1}$ and $\omega_{2}$, out of resonance with transitions
$g\leftrightarrow i$ and $e\leftrightarrow i$, respectively. Starting with the
engineering of the selective JC interactions, we write the Hamiltonian%

\[
H=\lambda\sigma_{ig}a\operatorname*{e}\nolimits^{-i\Delta t}+\Omega_{1}%
\sigma_{ig}\operatorname*{e}\nolimits^{i\Delta_{1}t}+\Omega_{2}\sigma
_{ie}\operatorname*{e}\nolimits^{-i\Delta_{2}t}+H.c.,
\]
where $\sigma_{rs}=\left\vert r\right\rangle \left\langle s\right\vert $, $r$
and $s$ labelling the atomic states involved, and the detunings are defined by
$\Delta=\omega-\omega_{ig}$, $\Delta_{1}=\omega_{ig}-\omega_{1}$, and
$\Delta_{2}=\omega_{2}-\omega_{ie}$, with $\omega_{i\ell}=\omega_{i}%
-\omega_{\ell}$ ($\ell=g,e$). Under the conditions $\sqrt{\bar{n}+1}\lambda
\ll\Delta$ and $\Omega_{j}\ll$ $\Delta_{j}$ ($j=1,2$), we readily derive, in
the RWA, the effective interaction \cite{James}
\begin{align*}
H_{eff}  &  =\left(  \xi a^{\dag}a-\varpi_{g}\right)  \sigma_{gg}+\varpi
_{e}\sigma_{ee}\\
&  +\left(  \zeta a^{\dagger}\operatorname*{e}\nolimits^{i\delta t}\sigma
_{ge}+H.c.\right)  ,
\end{align*}
where $\varpi_{g}=\left\vert \Omega_{1}\right\vert ^{2}/\Delta_{1}$ and
$\varpi_{e}=\left\vert \Omega_{2}\right\vert ^{2}/\Delta_{2}$ stand for
frequency-level shifts due to the action of the classical fields, whereas the
strengths $\xi=\left\vert \lambda\right\vert ^{2}/\Delta$ and $\zeta
=\lambda^{\ast}\Omega_{2}\left(  \Delta^{-1}+\Delta_{2}^{-1}\right)  /2$ stand
respectively for off- and on-resonant atom-field couplings to be used to
engineer the required selective interactions; finally,\textbf{ }$\delta
=\Delta-\Delta_{2}$\textbf{ }refers to a convenient detuning to be specified
in the following lines. To get selectivity, we first perform the unitary
transformation $U=\exp\left\{  -i\left[  \left(  \xi a^{\dag}a+\varpi
_{g}\right)  \sigma_{gg}+\varpi_{e}\sigma_{ee}\right]  t\right\}  $, which
takes $H_{eff}$ into the form%
\[
V_{eff}=%
{\textstyle\sum\nolimits_{n=1}^{\infty}}
\zeta_{n}\left\vert n+1\right\rangle \left\langle n\right\vert \sigma
_{ge}\operatorname*{e}\nolimits^{i\phi_{n}t}+H.c.\text{.}%
\]
with $\zeta_{n}=\sqrt{n+1}\zeta$ and $\phi_{n}=\left(  n+1\right)  \xi
+\delta-\varpi_{g}-\varpi_{e}$. Next, under the strongly off-resonant regime
$\xi\gg\sqrt{k+2}\left\vert \zeta\right\vert $ and the condition%

\[
\phi_{k}=0\text{,}%
\]
which is easily satisfied by imposing $\left(  k+1\right)  \xi$ $=$
$\varpi_{g}$ $\gg\delta=\varpi_{e}$, such that $\left\vert \Omega
_{1}\right\vert =\sqrt{\left(  k+1\right)  \Delta_{1}/\Delta}\left\vert
\lambda\right\vert \gg\sqrt{\Delta_{1}/\Delta_{2}}\left\vert \Omega
_{2}\right\vert $, we readily eliminate, in the RWA, all the terms
proportional to $\zeta_{n}=\sqrt{n+1}\zeta$ summed in $V_{eff}$, except when
$n=k$, bringing about the selective interaction%
\begin{equation}
\mathcal{H}=\left(  \zeta_{k}\left\vert k+1\right\rangle \left\langle
k\right\vert \sigma_{ge}+H.c.\right)  , \label{Eq13}%
\end{equation}
producing the desired selective $g\leftrightarrow e$ transition within the
Fock subspace $\left\{  \left\vert k\right\rangle ,\left\vert k+1\right\rangle
\right\}  $.

Next, following the reasoning in Ref. \cite{ABR,BR} for atomic reservoir
engineering, it is easily shown that the Lindblad structure of the
superoperator emerging from $\mathcal{H}$ does not rely on the weak-coupling
regime $\zeta_{k}\tau\ll1$, owing to the selective nature of this interaction.
With the atoms prepared in the excited state and the subspace $\left\{
\left\vert k\right\rangle ,\left\vert k+1\right\rangle \right\}  $ randomly
selected to be $\left\{  \left\vert 0\right\rangle ,\left\vert 1\right\rangle
\right\}  $ and $\left\{  \left\vert 1\right\rangle ,\left\vert 2\right\rangle
\right\}  $, which is achieved by an appropriate adjustment of parameters, we
readily obtain the selected Liouvillian (as well as from the inevitable
natural Liouvillian $\mathcal{L}\rho$)
\begin{equation}
\dot{\rho}=%
{\textstyle\sum\limits_{k=0}^{1}}
\frac{\Gamma_{k}}{2}\left(  2\left\vert k+1\right\rangle \left\langle
k\right\vert \rho\left\vert k\right\rangle \left\langle k+1\right\vert
\mathbf{-}\left\vert k\right\rangle \left\langle k\right\vert \rho
-\rho\left\vert k\right\rangle \left\langle k\right\vert \right)
+\mathcal{L}\rho, \label{Eq14}%
\end{equation}
with $\Gamma_{k}=r(\zeta_{k}\tau)^{2}$, which is somewhat different from that
in Eq. (\ref{Eq7}). In fact, the above equation does not contain the crossed
terms coupling together the two selected subspaces $\left\{  \left\vert
0\right\rangle ,\left\vert 1\right\rangle \right\}  $ and $\left\{  \left\vert
1\right\rangle ,\left\vert 2\right\rangle \right\}  $, that naturally arise in
Eq. (\ref{Eq7}). Evidently, we can build other $ub$ Liouvillians by selecting
additional subspaces beyond those arising from $k=0$ and $k=1$. As a final
remark, before applying the engineered Liouvillian (\ref{Eq14}) to produce
steady Fock states, we observe that this Liouvillian does not emerge from an
$ub$ Hamiltonian, as did the previous one given by Eq. (\ref{Eq7}). Its
derivation relies, instead, on the engineered selective interactions given by
Eq. (\ref{Eq13}).

To estimate the range of validity of the parameters leading to Hamiltonian
(\ref{Eq13}), we start by choosing $\Delta=\Delta_{1}=$ $(1+10^{-2}%
)\times\Delta_{2}=10\sqrt{k+1}\left\vert \lambda\right\vert $, such that
$\left\vert \Omega_{1}\right\vert =10\times\left\vert \Omega_{2}\right\vert
=\sqrt{k+1}\left\vert \lambda\right\vert $, $\zeta_{k}=10^{-2}\sqrt
{k+1}\left\vert \lambda\right\vert $, and $\tau=1/r=10^{2}/\sqrt{\max
(k)+1}\left\vert \lambda\right\vert $, so that $\zeta_{\max(k)}\tau=1$, and
$\Gamma_{k}=\left\{  \left(  k+1\right)  \left\vert \lambda\right\vert
/\left[  \max(k)+1\right]  ^{3/2}\right\}  \times10^{-2}$. Assuming typical
$\lambda\sim5\times10^{5}$Hz, $\gamma\sim10$Hz \cite{RMP} and $\bar{n}=0.05$,
in Fig. 6(a), we plot the evolution of the fidelity $\mathcal{F}_{2}(t)=$
$\operatorname*{Tr}\left\vert 2\right\rangle \left\langle 2\right\vert
\rho(t)$ against $\gamma t$, setting $\tau=1/r=\sqrt{2}\times10^{-4}$s,
$\Gamma_{0}=$ $176\gamma$ and $\Gamma_{1}=$ $2\Gamma_{0}$. We obtain a
fidelity around $0.95$. Using the same typical parameters, in Fig. 6(b) we
display the evolution of the fidelity $\mathcal{F}_{3}(t)$ against $\gamma t$,
with $\tau=1/r=2\times10^{-4}/\sqrt{3}$s, $\Gamma_{0}=$ $96\gamma$,
$\Gamma_{1}=$ $2\Gamma_{0}$, and $\Gamma_{2}=$ $3\Gamma_{0}$, reaching a
fidelity around $0.94$. In the inset in Figs. 6(a and b), we plot Mandel's $Q$
factor, which tends to $-0.98$ and $-0.97$, respectively, showing that we
indeed approach a Fock state in both cases.
\begin{figure}
\includegraphics[width=1.0\columnwidth]{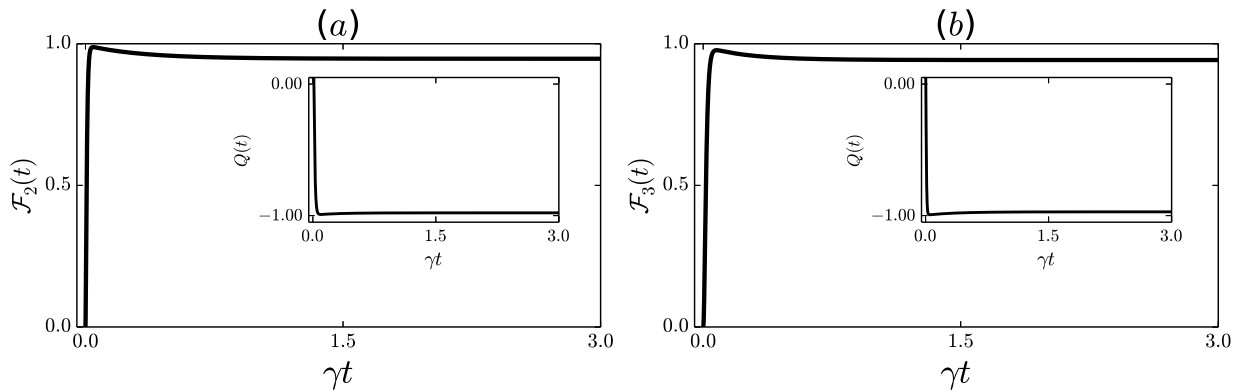}%
\caption{(a) Evolution of the fidelity $\mathcal{F}_{2}(t)=$ $\operatorname*{Tr}%
\left\vert 2\right\rangle \left\langle 2\right\vert \rho(t)$, governed by Eq.
(\ref{Eq14}), showing that a steady Fock state $\left\vert 2\right\rangle $ is
reached with a significant fidelity, around $0.95$, from a thermal state. The
evolution of Mandel's $Q$ factor, shown in the inset, leads to a value around
$-0.98$, indicating that the final steady state indeed approaches a Fock
state. (b) Evolution of the fidelity $\mathcal{F}_{3}(t)=$ $\operatorname*{Tr}%
\left\vert 3\right\rangle \left\langle 3\right\vert \rho(t)$, governed by Eq.
(\ref{Eq14}), showing that a steady Fock state $\left\vert 3\right\rangle $ is
reached with a significant fidelity, around $0.94$, from a thermal state. The
evolution of Mandel's $Q$ factor, shown in the inset, leads to a value around
$-0.97$, indicating that the final steady state indeed approaches a Fock state.}%

\end{figure}
\section{Conclusions}

We have thus presented a technique to engineer $ub$ and sliced atom-field
interactions in cavity QED. We started by engineering these interactions
confined to the Fock subspaces $\left\{  \left\vert 0\right\rangle ,\left\vert
1\right\rangle ,\left\vert 2\right\rangle \right\}  $ and $\left\{  \left\vert
M\right\rangle ,\left\vert M+1\right\rangle ,\left\vert M+2\right\rangle
\right\}  $, respectively, using a four-level atomic configuration, a cavity
mode and a laser field. Next, with the same cavity mode and laser field but
with a five-level atomic configuration, we showed how to engineer interactions
confined to the Fock subspaces ranging from $\left\vert 0\right\rangle $ to
$\left\vert 3\right\rangle $ and $\left\vert M\right\rangle $ to $\left\vert
M+3\right\rangle $. The validity of these engineered interactions was
confirmed by comparing the evolution of the cavity mode population described
by them with those described by the full Hamiltonians from which they were
derived. All the numerical simulations in this work were performed using QuTIP
\cite{QuTIP}. We have also indicated how to generate, through a six-level
atomic configuration, interactions confined to the Fock subspaces ranging from
$\left\vert 0\right\rangle $ to $\left\vert 3\right\rangle $ and $\left\vert
M\right\rangle $ to $\left\vert M+4\right\rangle $.

Next, turning to the applications of the derived $ub$ Hamiltonians confined to
a subspace of states ranging from $\left\vert 0\right\rangle $ up to
$\left\vert N\right\rangle $, we presented a method, relying on atomic
reservoir engineering \cite{ABR,BR}, for the protection of the higher energy
state $\left\vert N\right\rangle $. Focusing on the $ub$ Hamiltonian confined
to the Fock subspaces $\left\{  \left\vert 0\right\rangle ,\left\vert
1\right\rangle ,\left\vert 2\right\rangle ,\left\vert 3\right\rangle \right\}
$, we presented the evolution of the fidelity $\mathcal{F}_{3}(t)=$
$\operatorname*{Tr}\left\vert 3\right\rangle \left\langle 3\right\vert
\rho(t)$ and Mandel's $Q$ factor, showing that the field reaches the steady
state that indeed approaches the Fock state $\left\vert 3\right\rangle $.

We then present a strategy ---following the reasoning in Ref. \cite{Wilson},
where engineered atomic reservoirs \cite{ABR,BR} are also assumed--- to
generate $ub$ Liouvillians, which does not rely on the engineering of a
specific $ub$ Hamiltonian. The engineered Liouvillian is then used to generate
and protect states that approach closely the Fock states $\left\vert
2\right\rangle $ and $\left\vert 3\right\rangle $.

The two schemes outlined above for the production of steady Fock states have
approximately the same effectiveness and difficulty of practical
implementation. The engineering of $ub$ Hamiltonians and Liouvillians becomes
increasingly difficult as we increase the size of the $ub$ subspace. While the
engineering of $ub$ interactions requires an increasing number of specific
atomic levels and laser beams, the engineering of $ub$ Liouvillians demands
the random selection of an increasing number of the subspaces $\left\{
\left\vert k\right\rangle ,\left\vert k+1\right\rangle \right\}  $.

We finally observe that a number of distinct applications of the $ub$ and
sliced interactions we have engineered above can be implemented, such as the
engineering of non-classical states, quantum logical gates, teleportation of
superpositions of $N>2$, etc. Basically, what we have done in this
contribution can be taken forward in the context of what have been called
quantum-scissors for optical state truncation \cite{Others}. As already
stressed above, the technique here presented can be directly applicable, among
others platforms, to circuit QED \cite{CQED}.

The authors acknowledge financial support from PRP/USP within the Research
Support Center Initiative (NAP Q-NANO) and FAPESP, CNPQ and CAPES, Brazilian agencies.

\end{document}